\documentclass[%
superscriptaddress,
amsmath,amssymb,
aps,
prfluids,
longbibliography,
11pt
]{revtex4-2}

\usepackage{amssymb}
\usepackage{amsthm}
\usepackage{graphicx}
\usepackage{amsmath}
\usepackage{amsfonts}
\usepackage{color}
\usepackage{centernot}
\usepackage{mathtools}
\usepackage{stmaryrd}

\newtheorem{theorem}{Theorem}

\begin{document}
\title{Shell model intermittency is the hidden self-similarity}
\author{Alexei A. Mailybaev} 
\affiliation{Instituto de Matem\'atica Pura e Aplicada -- IMPA, Rio de Janeiro, Brazil}
\email{alexei@impa.br}

\begin{abstract}
We show that the intermittent dynamics observed in the inertial interval of Sabra shell model of turbulence can be rigorously related to the property of scaling self-similarity. In this connection, the space-time scaling symmetries (like in the K41 theory) are replaced by the new hidden scaling symmetry, which is an exact symmetry of inviscid dynamics represented in special rescaled coordinates and times. We derive the formulas expressing anomalous scaling exponents in terms of Perron--Frobenius eigenvalues of linear operators based on the self-similar statistics. Theoretical conclusions are verified by extensive numerical simulations.
\end{abstract}
\maketitle

\section{Introduction} 

Small-scale intermittency remains one of major open problems in the theory of developed turbulence. This intermittency is featured by alternating regions of intense turbulent and extended laminar behaviors becoming more pronounced at smaller scales. Quantitative observables usually associated with the intermittency are the structure functions, e.g. $S_p(\ell) = \langle \delta v_\ell^p\rangle$ defined as the averaged moments of velocity differences $\delta v_\ell$ at scale $\ell$~\cite{frisch1999turbulence}. In the Kolmogorov’s theory of 1941, the self-similarity assumption leads to the power-law scaling $S_p(\ell) \propto \ell^{\zeta_p}$ with the linear dependence of the scaling exponents $\zeta_p = p/3$ on the order $p$. This is, however, in contradiction with experimental and numerical observations indicating that the dependence of $\zeta_p$ on the order $p$ is nonlinear, also called the anomalous scaling. Anomalous scaling implies that all space-time scaling symmetries are broken in the statistics of developed turbulence, leading to a much higher complexity of the flow. Such intermittent flows may be considered in the context of multifractal approach~\cite{frisch1999turbulence}. We recall that the discussed properties refer to the scales in the so-called inertial interval, where both forcing and viscous terms are negligible. 

Shell models were introduced as toy models with the aim of revealing mechanisms that lead to intermittency. Among numerous shell models considered up to now, the most popular are probably the GOY model~\cite{gledzer1973system,ohkitani1989temporal} and its modification named the Sabra model~\cite{l1998improved}. These models demonstrated principal features of the Navier--Stokes intermittency, including the anomalous scaling of structure functions with the exponents $\zeta_p$ rather close to their values in the full Navier--Stokes system. 
Multiple efforts were undertaken for understanding the shell model intermittency theoretically~\cite{biferale2003shell}, e.g., with the multifractal approach~\cite{kadanoff1995scaling} or computing scaling exponents~\cite{l2000analytic,benzi2003intermittency} with the use of fusion rules~\cite{eyink1993lagrangian,l1996fusion}. However, these derivations are not free from phenomenological assumptions. Therefore, theoretical explanation of intermittency from first principles is still missing for shell models.

In this work we show that intermittency in the Sabra shell model can be derived from the hidden scaling symmetry reported in~\cite{mailybaev2021hidden,mailybaev2020hidden}. We demonstrate that this new symmetry arises, when the inviscid equations of motion are represented in a rescaled form with an intrinsic (solution-dependent) time. We show that the property of hidden scale self-similarity in the inertial interval yields the anomalous scaling of structure functions. Thus, it is the hidden rather than usual self-similarity that controls the statistics of developed turbulence. Following a general group-theoretical approach of~\cite{mailybaev2020hidden}, we express the scaling exponents $\zeta_p$ in terms of Perron--Frobenius eigenvalues of certain operators derived from the  self-similar statistics. The suggested new formalism develops and provides the the first-principles explanation to earlier ideas related to the concept of multipliers, first appeared in the famous work Kolmogorov in 1962~\cite{kolmogorov1962refinement}. These originally phenomenological ideas were inspired by the theory of multiplicative stochastic processes and later discussed in relation to intermittency for shell models~\cite{benzi1993intermittency,eyink2003gibbsian,PhysRevX.11.021063}. 

Our formalism naturally unifies the ideas of Kolmogorov with those of Parisi-Frisch multifractal theory of turbulence~\cite{frisch1985singularity,frisch1999turbulence}. The latter proposes that all space-time scaling symmetries are restored in the inertial interval of developed turbulence, but each symmetry within a corresponding fractal subset of space-time depending on a scaling exponent $h$. We prove that our construction fuses the one-parameter family of space-time scaling symmetries (depending on $h$) into the single hidden symmetry, therefore, reducing the Parisi--Frisch argument to the restoration of the hidden symmetry alone and in the usual sense. Notice that the existence of such kind of symmetry in intermittent turbulence was anticipated by She\&Leveque in 1994~\cite{she1994universal}, where the
authors wrote: “We believe that this relation is a consequence of some hidden (statistical) symmetries in the solution of the Navier-Stokes equations.”

We start with the description of the Sabra shell model in Section~\ref{sec2}. Section~\ref{sec3} introduces the hidden scaling symmetry. Structure functions are expressed in terms of rescaled variables and times in Section~\ref{sec4}, and these expressions are used in Section~\ref{sec5} for deriving scaling exponents in terms of Perron--Frobenius eigenvalues. Section~\ref{sec6} provides the detailed verification of obtained results by numerical simulations. In Section~\ref{sec7}, we discuss the results and their implications for the Navier--Stokes turbulence. The Appendix contains some technical derivations.

\section{Model} \label{sec2}

The shell model of turbulence mimics the Navier--Stokes flow using a geometric sequence of wavenumbers $k_n = \lambda^n$ with $\lambda = 2$ and corresponding complex velocities $u_n \in \mathbb{C}$ indexed by integer shell numbers $n$. Spatial scales are defined as $\ell_n = 1/k_n$, i.e., larger shell numbers correspond to smaller scales. Equations of the Sabra model~\cite{l1998improved} are formulated in the dimensionless form as
    \begin{equation}
    \frac{du_n}{dt} = ik_n\left(2u_{n+2}u_{n+1}^*
	-\frac{u_{n+1}u_{n-1}^*}{2} 
	+\frac{u_{n-1}u_{n-2}}{4} \right)
	-\mathrm{Re}^{-1} k_n^2u_n,\quad 
	n \ge 1.
	\label{eq1a}
    \end{equation}
The right-hand side of these equations contains the quadratic nonlinear term and the viscous term multiplied by the inverse of the Reynolds number $\mathrm{Re} > 0$.  Shell model (\ref{eq1a}) possesses two inviscid invariants interpreted as the energy $E = \frac{1}{2}\sum_n |u_n|^2$ and helicity $H = \sum_n(-1)^nk_n|u_n|^2$~\cite{l1998improved}. It is convenient to set 
    \begin{equation}
    u_0(t) \equiv 1
    \label{eq2b}
    \end{equation}
as the boundary (forcing) condition with $u_n(t) \equiv 0$ for negative $n$.

We consider the regime of developed turbulence corresponding to very large Reynolds numbers $\mathrm{Re} \gg 1$. In the statistically stationary state, the inertial interval is defined as the range of shells $n$ such that
    \begin{equation}
    k_0 \ll k_n \ll K.
    \label{eq_IR}
    \end{equation}
This interval is distant from the large-scale forcing condition (\ref{eq2b}) at $k_0 = 1$ (forcing range) and from large wavenumbers of order $K$, for which the viscosity starts playing the role (viscous range). The Kolmogorov theory (K41) estimates 
$K \sim \mathrm{Re}^{3/4}$~\cite{frisch1999turbulence}. Considering shells with wavenumbers $k_n \ll K$, i.e., the inviscid scales of forcing range and inertial interval, we can neglect the viscosity by setting $\mathrm{Re}^{-1} = 0$ in system (\ref{eq1a}). This yields the equation
    \begin{equation}
    \frac{du_n}{dt} = ik_n\left(2u_{n+2}u_{n+1}^*
	-\frac{u_{n+1}u_{n-1}^*}{2} 
	+\frac{u_{n-1}u_{n-2}}{4} \right),
	\label{eq1Euler}
    \end{equation}
which mimics the Euler system for ideal fluid. One can see that Eq.~(\ref{eq1Euler}) is invariant with respect to  a family of space-time scalings
    \begin{equation}
    t,\ u_n \mapsto \lambda^{1-h}t,\ \lambda^h u_{n+1}.
    \label{eq_S1}
    \end{equation}
Here the exponent $h \in \mathbb{R}$ defines an arbitrary temporal scaling factor $\lambda^{1-h}$, and the shift of all shells by the unity induces the change of spatial scale $k_{n+1} = \lambda k_n$ with $\lambda = 2$. Transformations (\ref{eq_S1}) mimic space-time scaling symmetries of the Euler equations~\cite{frisch1999turbulence}.

Structure functions are traditional observables for the analysis of intermittency in stationary developed  turbulence. For the shell model, they are introduced as
    \begin{equation}
    S_p(k_n) 
    = \left\langle|u_n|^p\right\rangle_t,
    \label{eq1_Sp}
    \end{equation}
where $\langle \cdot \rangle_t$ denotes the temporal average and $p \ge 0$ is the order of velocity moment. 
Numerical simulations~\cite{l1998improved} demonstrate an accurate power-law scaling 
    \begin{equation}
    S_p(k_n) \propto k_n^{-\zeta_p}
    \label{eq1_Sp2}
    \end{equation}
in the inertial range. The nonlinear dependence of exponents $\zeta_p$ on $p$ is a distinctive feature of the intermittency, because the K41 theory based on the scale invariance predicts the linear dependence $\zeta_p = p/3$. Deviations $\zeta_p-p/3$ of the actual exponents from the K41 theory are called the \textit{anomalous corrections}. These corrections are relatively small, vanishing  for $\zeta_0 = 0$ and $\zeta_3 = 1$ but getting large for high-order moments. Power laws (\ref{eq1_Sp2}) with anomalous corrections are the main focus for our work. 

\section{Hidden scaling symmetry}\label{sec3}

Let us fix some reference shell number $m$ and introduce the corresponding momentary turn-over time as
    \begin{equation}
    T_m(t) = \bigg(\sum_{j < m} k_j^2|u_j(t)|^2\bigg)^{-1/2}.
    \label{eq2_T}
    \end{equation}
Expression in the parentheses is an analogue of enstrophy for the shells $j = 0,\ldots,m-1$.
Then, we define the rescaled variables $U_N$ as functions of the intrinsic time $\tau$ implicitly as~\cite{mailybaev2021hidden}
    \begin{equation}
    d\tau = \frac{d t}{T_m( t)},\quad
    U_N = ik_mT_m(t)u_{N+m}(t)
    \label{eq2}
    \end{equation}
with the initial condition $\tau = 0$ at $t = 0$.
Writing Eq.~(\ref{eq1Euler}) for the inviscid dynamics in terms of new variables (\ref{eq2}), we have (see Appendix~\ref{secA1} for the derivation)
	\begin{equation}
	\label{eqZ_1}
	\frac{dU_N}{d\tau} 
	= -k_N B_N+U_N\sum_{J = 1-m}^{-1}
	k_J^3\mathrm{Re}(U_J^*B_J),
	\end{equation}
where $\mathrm{Re}(\cdot)$ denotes the real part and
	\begin{equation}
	\label{eqZ_1b}
	B_N = 
	2U_{N+2}U_{N+1}^*
	-\frac{U_{N+1}U_{N-1}^*}{2} 
	-\frac{U_{N-1}U_{N-2}}{4}.
	\end{equation}
Notice that the sum in (\ref{eqZ_1}) represents the derivative $dT_m/dt$.

Using the K41 estimate $u_j \sim k_j^{-1/3}$ in Eqs.~(\ref{eq2_T}), (\ref{eq2}) and (\ref{eqZ_1b}) yields 
	\begin{equation}
	\label{eqK41}
	k_j^2|u_j|^2 \sim k_j^{4/3},\quad
	T_m \sim k_m^{-2/3},\quad
	U_N \sim k_N^{-1/3},\quad
	k_J^3U_J^*B_J \sim k_J^2.
	\end{equation}
The intermittency adds small corrections to the exponents. We see that the sum in expression (\ref{eqZ_1}) is dominated by $J$ close to $0$. Therefore, considering the reference shell $m$ with $k_m \gg k_0$ from the inertial interval (\ref{eq_IR}), one can write Eq.~(\ref{eqZ_1}) formally as 
	\begin{equation}
	\label{eqZ_1_sym}
	\frac{dU_N}{d\tau} 
	= -k_N B_N+U_N\sum_{J < 0}
	k_J^3\mathrm{Re}(U_J^*B_J),
	\end{equation}
i.e., ignoring the cutoff of the sum at $J = 1-m$.
  
The key observation is that the resulting system (\ref{eqZ_1_sym}) and (\ref{eqZ_1b}) does not depend explicitly on $m$. Hence, the change of the reference shell $m$ defines a symmetry transformation for this system. This transformation can be written in an explicit form. For example, the increase of reference shell by unity yields the new intrinsic time and rescaled variables (denoted by hats) for $\hat{m} = m+1$ as~\cite{mailybaev2021hidden}
	\begin{equation}
	\label{eqRV_plus}
	d\hat{\tau} = \sqrt{1+|U_0|^2} \,d\tau,\quad
	\hat{U}_N = \frac{2U_{N+1}}{\sqrt{1+|U_0|^2}}.
	\end{equation}
Transformations (\ref{eqRV_plus}) represent the simultaneous state-dependent change of time and nonlinear change of variables.
The invariance of equations (\ref{eqZ_1_sym}) and (\ref{eqZ_1b}) with respect to transformation (\ref{eqRV_plus}) can be verified by a direct substitution; see the Appendix~\ref{secA_2}. 

We conclude that transformations (\ref{eqRV_plus}) define the \textit{hidden scaling symmetry} of system (\ref{eqZ_1_sym}) and (\ref{eqZ_1b}). It is a weaker symmetry: the hidden scale invariance can be restored in a statistical solution despite all original symmetries (\ref{eq_S1}) are broken~\cite{mailybaev2020hidden,mailybaev2021solvable}. We remark that the particular choice of expression (\ref{eq2_T}) for turn-over times is not unique, and equivalent formulations of the hidden symmetry can be obtained with other definitions~\cite{mailybaev2020hidden}. 

Let us now relate the hidden symmetry with the original space-time scaling symmetries (\ref{eq_S1}). Specifically, let us consider the scaled shell velocities and times as 
	\begin{equation}
	\label{eqSF1}
	\hat{u}_n(\hat{t}) = \lambda^h u_{n+1}(t), \quad \hat{t} = \lambda^{1-h} t,
	\end{equation}
which are obtained by the scaling (\ref{eq_S1}) for an arbitrary $h \in \mathbb{R}$. One can easily verify that velocities (\ref{eqSF1}) satisfy the inviscid equations of motion (\ref{eq1Euler}) if the original velocities $u_n(t)$ do. By substituting (\ref{eqSF1}) into expressions (\ref{eq2_T})--(\ref{eq2}), one can show (see Appendix~\ref{secA_2b}) that the corresponding  rescaled velocities $\hat{U}_N(\hat{\tau})$ are given by the hidden symmetry relations (\ref{eqRV_plus}). Although it appears that the hidden symmetry is a rescaled version of the original space-time scaling symmetry, the crucial difference is that the hidden symmetry does not depend on the exponent $h$. Hence, the rescaling procedure fuses the one-parameter family of space-time scaling symmetries into a single hidden symmetry. This fusion is not just a property of a particular model under consideration, but a general consequence of commutation relations in the symmetry group as shown in~\cite{mailybaev2020hidden}. 

The fusion of scaling symmetries into the hidden symmetry connects our formalism with the Parisi--Frisch multifractal theory of turbulence~\cite{frisch1985singularity,frisch1999turbulence}. The latter proposes that all space-time scaling symmetries are restored in the inertial interval of developed turbulence, but each symmetry within a corresponding fractal subset of space-time depending on $h$. The fusion property of our construction reduces the Parisi--Frisch argument to the restoration of the hidden symmetry alone and in the usual sense.
Indeed, we show below that the restoration of hidden scale invariance is both the case and the cause for the intermittent turbulent dynamics.

\section{Structure functions in terms of rescaled variables}\label{sec4}

Our next goal is to express the structure functions in terms of rescaled variables. For this purpose, we introduce the auxiliary real variables, which we call \textit{multipliers}, as
    \begin{equation}
	\sigma_N = \frac{T_{N+m}}{T_{N+m+1}},\quad
	N > -m.
    \label{eqAx_1mult}
    \end{equation}
Using definitions (\ref{eq2_T}) and (\ref{eq2}) in (\ref{eqAx_1mult}), one can see that 
    \begin{equation}
	\sigma_N =  \sqrt{\frac{\sum_{J \le N}k_J^2|U_J|^2}{\sum_{J < N}k_J^2|U_J|^2}}.
    \label{eqAx_1}
    \end{equation}
Clearly, 
    \begin{equation}
	\sigma_N \ge 1
    \label{eqAx_1MP}
    \end{equation}
for all multipliers. 
Notice that ratios of turn-over times (\ref{eqAx_1mult}) are analogous to the Kolmogorov multipliers defined as ratios of velocities, whose universality was discussed in several previous works \cite{benzi1993intermittency,eyink2003gibbsian,mailybaev2016spontaneously}. However, expressions (\ref{eqAx_1mult}) have the advantage of avoiding vanishing denominators. 
Notice that $k_J^2|U_J|^2 \sim k_J^{4/3}$ by the K41 estimate (\ref{eqK41}) with a small correction of exponent due to intermittency. Hence, the sums in (\ref{eqAx_1}) are dominated by $J$ close to $N$. 

Let us consider a statistically stationary state of developed turbulence. We assume the existence of a probability distribution with ergodic properties, which expresses averages with respect to the intrinsic time $\tau$. Specifically, let us introduce the vectors 
    \begin{equation}
    \boldsymbol{\sigma} 
    = \left(\sigma_0,\boldsymbol{\sigma}_-\right) \in \mathbb{R}^m,\quad
    \boldsymbol{\sigma}_- = (\sigma_{-1},\ldots,\sigma_{-m+1}) \in \mathbb{R}^{m-1}.
    \label{eqM_1}
    \end{equation}
We consider a probability measure $\mu(\boldsymbol{\sigma})$ in the space $\boldsymbol{\sigma} \in \mathbb{R}^m$ such that  
    \begin{equation}
    \langle \varphi \rangle_\tau = \int \varphi(\boldsymbol{\sigma})\,d\mu(\boldsymbol{\sigma})
    \label{eqAv_1}
    \end{equation}
for any continuous observable $\varphi(\boldsymbol{\sigma})$, where $\langle \cdot \rangle_{\tau}$ denotes the average with respect to time $\tau$. We denote by $\rho(\sigma_0|\boldsymbol{\sigma}_-)$ a conditional probability density defined in the standard way by the relation 
    \begin{equation}
    d\mu(\boldsymbol{\sigma}) = \rho(\sigma_0|\boldsymbol{\sigma}_-) \,d\sigma_0\, d\mu_-(\boldsymbol{\sigma}_-),
    \label{eqAv_1cond}
    \end{equation}
where $\mu_-(\boldsymbol{\sigma}_-)$ is a probability measure for the vector $\boldsymbol{\sigma}_- \in \mathbb{R}^{m-1}$. The following statement (see Appendix~\ref{secA_3} for the proof) expresses the structure functions in terms of the measure $\mu$.

\begin{theorem}\label{theorem1}
Structure functions (\ref{eq1_Sp}) at the reference shell number $m$ can be expressed as
    \begin{equation}
    S_p(k_m) 
    = \int (\sigma_0^2-1)^{p/2}d\mu_p(\boldsymbol{\sigma})
    \label{eqZ_6prop}
    \end{equation}
integrated with the measure 
    \begin{equation}
    d\mu_p(\boldsymbol{\sigma}) = 
    k_m^{-p}\left\langle T_m^{-1} \right\rangle_t 
    \left(\prod_{J = 1-m}^{-1}\sigma_J^{p-1}\right) d\mu(\boldsymbol{\sigma}).
    \label{eqZ_7prop}
    \end{equation}
Under the unit increase of reference shell, $\hat{m} = m+1$, the new multipliers are expressed as $\hat{\sigma}_N = \sigma_{N+1}$, and
corresponding new measure $\hat{\mu}_p(\hat{\boldsymbol{\sigma}})$ satisfies the iterative relation
    \begin{equation}
    d\hat{\mu}_p(\hat{\boldsymbol{\sigma}}) = \left(\frac{\hat{\sigma}_{-1}}{\lambda}\right)^p
    \hat{\rho}(\hat{\sigma_0}|\hat{\boldsymbol{\sigma}}_-)\,
    d\hat{\sigma}_0\, d\mu_p(\hat{\boldsymbol{\sigma}}_-),
    \label{eqZ_8prop}
    \end{equation}
where 
    \begin{equation}
	\hat{\boldsymbol{\sigma}} = (\hat{\sigma}_0,\hat{\boldsymbol{\sigma}}_-) \in \mathbb{R}^{\hat{m}},\quad 
	\hat{\boldsymbol{\sigma}}_- = (\hat{\sigma}_{-1},\hat{\sigma}_{-2},\ldots,\hat{\sigma}_{-\hat{m}+1}) \in \mathbb{R}^{\hat{m}-1}, 
    \label{eqZ_8prop_B}
    \end{equation}
and the conditional probability density $\hat{\rho}(\hat{\sigma}_0|\hat{\boldsymbol{\sigma}}_-)$ corresponds to the new reference shell $\hat{m}$.
\end{theorem}

\section{Anomalous exponents as Perron--Frobenius eigenvalues} \label{sec5}

In this section, we study the dependence of statistics on the reference shell $m$ and, for the clarity, adopt slightly different notations. Let us write the iterative relation (\ref{eqZ_8prop}) in Theorem~\ref{theorem1} as
    \begin{equation}
    d\mu_p^{(m+1)}(\sigma_0,\sigma_{-1},\ldots,\sigma_{-m}) 
    = \left( \frac{\sigma_{-1}}{\lambda}\right)^p
    \rho^{(m+1)}(\sigma_0|\sigma_{-1},\ldots,\sigma_{-m})
    \,d\sigma_0
    \,d\mu_p^{(m)}(\sigma_{-1},\ldots,\sigma_{-m}),
    \label{eqPF_4}
    \end{equation}
where we specified the reference shells $m$ and $\hat{m} = m+1$ explicitly in the superscripts and dropped the hats. This iterative relation can be written in a short form as
    \begin{equation}
    \mu_p^{(m+1)} = \mathcal{L}_p^{(m+1)} [\mu_p^{(m)}] ,
    \label{eqPF_3}
    \end{equation}
where $\mathcal{L}_p^{(m+1)}$ is defined by (\ref{eqPF_4}) as a linear operator acting on measures $\mu_p^{(m)}$.
By induction, for any $m$, we write 
    \begin{equation}
    \mu_p^{(m)} = \mathcal{L}_p^{(m)} \circ \mathcal{L}_p^{(m-1)} \circ \cdots \circ \mathcal{L}_p^{(2)} [\mu_p^{(1)}].
    \label{eqPF_5}
    \end{equation}
For $m = 1$, expressions (\ref{eq2}) and (\ref{eq2bb}), (\ref{eqAx_1y}) from Appendix~\ref{secA_3} yield $T_1(t) \equiv 1$, $|U_0| = k_1 |u_1|$ and $\sigma_0 = \sqrt{1+k_1^2|u_1|^2}$. Thus, the measure $\mu_p^{(1)}$ is found from (\ref{eqZ_7prop}) as
    \begin{equation}
	d\mu_p^{(1)}(\sigma_0) = k_1^{-p} \, d\mu^{(1)}(\sigma_0),
    \label{eqPF_6}
    \end{equation}
where $\mu^{(1)}(\sigma_0)$ is the probability measure for the statistical distribution of $\sigma_0 = \sqrt{1+k_1^2|u_1|^2}$ with respect to the original time $\tau_1 = t$.

We define the \textit{statistical hidden self-similarity} as the invariance of a probability distribution with respect to the hidden symmetry transformations (\ref{eqRV_plus}). For the developed turbulence, this self-similarity may be expected in the inertial interval, where equations of motion are invariant with respect to the hidden symmetry. Recall that transformations (\ref{eqRV_plus}) are equivalent to a change of the reference shell $m$. Hence, the hidden self-similarity implies that the conditional probability density $\rho^{(m)}$ does not depend on the reference shells $m$ in the inertial interval:
    \begin{equation}
	\rho^{(m)}(\sigma_0|\sigma_{-1},\ldots,\sigma_{1-m})  \approx \rho^{\infty}(\sigma_0|\sigma_{-1},\sigma_{-2},\ldots),
    \label{eqHSS_1}
    \end{equation}
where we use $\infty$ in the superscript for denoting self-similar quantities.
We also require that correlations between $\sigma_0$ and $\sigma_J$ with $J < 0$ are local, i.e., the dependence of $\rho^{\infty}$ on variables $\sigma_J$ decays for large negative $J$.

The self-similarity assumption (\ref{eqHSS_1}) implies the invariance property of the linear operator from (\ref{eqPF_4}) and (\ref{eqPF_3}). Thus, we have
    \begin{equation}
	\mathcal{L}_p^{(m)}  \approx \mathcal{L}_p^{\infty}   
    \label{eqHSS_2}
    \end{equation}
for reference shells $m$ in the inertial interval. Here $\mathcal{L}_p^{\infty}$ is a linear operator defined by (\ref{eqPF_4}) and (\ref{eqHSS_1}) as
    \begin{equation}
    \hat\mu = \mathcal{L}_p^{\infty} [\mu],\quad
    d\hat\mu(\sigma_0,\sigma_{-1},\sigma_{-2},\ldots) = 
    \left( \frac{\sigma_{-1}}{\lambda}\right)^p
    \rho^{\infty}(\sigma_0|\sigma_{-1},\sigma_{-2},\ldots)
    \,d\sigma_0
    \,d\mu(\sigma_{-1},\sigma_{-2},\ldots),
    \label{eqHSS_3}
    \end{equation}
which acts on measures $\mu(\sigma_0,\sigma_{-1},\sigma_{-2},\ldots)$ in the infinite-dimensional space. 

Since the operator $\mathcal{L}_p^{\infty}$ is positive (mapping positive-valued measures to positive-valued measures), it has a unique dominant eigenvalue $R_p$ with the corresponding eigenvector (measure) $\mu_p^{\infty}$:
    \begin{equation}
    \mathcal{L}_p^{\infty} [\mu_p^{\infty}] = R_p \,\mu_p^{\infty};
    \label{eqHSS_4}
    \end{equation} 
see the Perron--Frobenius theorem for positive matrices~\cite[Ch.~16]{lax2007linear} and the Krein--Rutman theorem~\cite[\S 19.5]{deimling2010nonlinear} for positive operators with proper assumptions of compactness. The Perron--Frobenius eigenvalue $R_p$ is real positive and larger than absolute values of all remaining eigenvalues. 
Hence, for a generic measure (\ref{eqPF_6}) in the forcing range and large $m$ in the inertial interval, expressions (\ref{eqPF_5}) and (\ref{eqHSS_2}) yield the asymptotic form of measure $\mu_p^{(m)}$ as
    \begin{equation}
   \mu_p^{(m)} \approx c_p R_p^m \mu_p^{\infty},
    \label{eqHSS_5}
    \end{equation}
where the coefficient $c_p$ depends on statistical properties in the forcing range. 

Substituting (\ref{eqHSS_5}) into expression (\ref{eqZ_6prop}) yields
    \begin{equation}
    S_p(k_m) \approx C_p R_p^m \propto k_m^{-\zeta_p}
    \label{eqHSS_6}
    \end{equation}
with the scale-independent coefficient 
    \begin{equation}
    C_p = c_p\int (\sigma_0^2-1)^{p/2}
    \,d\mu_p^{\infty}(\sigma_0,\sigma_{-1},\ldots)
    \label{eqHSS_7}
    \end{equation}
and the exponent
    \begin{equation}
	\zeta_p = -\log_\lambda R_p.
    \label{eqHSS_8}
    \end{equation}
If the constant $C_p$ in (\ref{eqHSS_7}) is finite (neither vanishing nor infinite), then expression (\ref{eqHSS_8}) yields our main result: The statistical hidden self-similarity implies the power-law scaling of structure functions with the exponents (\ref{eqHSS_8}) given by Perron--Frobenius eigenvalues of linear operators (\ref{eqHSS_3}). From the form of these operators, one can argue~\cite{mailybaev2020hidden} that exponents (\ref{eqHSS_8}) can be and typically are anomalous, i.e., they depend nonlinearly on $p$; see e.g. \cite{mailybaev2021solvable} for an explicit calculation in a solvable intermittent shell model. 

We finally notice that, for the mathematical definition of asymptotic relations (\ref{eqHSS_1}), (\ref{eqHSS_2}) and (\ref{eqHSS_5}), one should consider the double limit: the first limit of large Reynolds number $\mathrm{Re} \to \infty$ and the subsequent limit of large wavenumber $k_m \to \infty$. 
Convergences may be understood in terms of the standard product topology in infinite-dimensional space $(\sigma_0,\sigma_{-1},\sigma_{-2},\ldots) \in \mathbb{R}^\infty$; see e.g. \cite{tao2011introduction}.

\section{Numerical tests}\label{sec6}

In this section, we use results of numerical simulations for the detailed confirmation of the hidden self-similarity and its relation with the intermittency. In these simulations, we used the total number of $n = 40$ shells and the Reynolds number $\mathrm{Re} = 10^{12}$. Equations (\ref{eq1a}) were simulated with high accuracy  in the large time interval $0 \le t \le 2000$ corresponding to the statistically stationary regime. Simultaneously intrinsic times (\ref{eq2}) were integrated for different reference shells. As initial conditions, we considered the K41 state $u_n(0) = e^{i\theta_n}k_n^{-1/3}$ with random phases and omitted the initial interval of length $\Delta t = 20$ for removing a transient behavior.

\begin{figure}[t]
\centering
\includegraphics[width=1.00\textwidth]{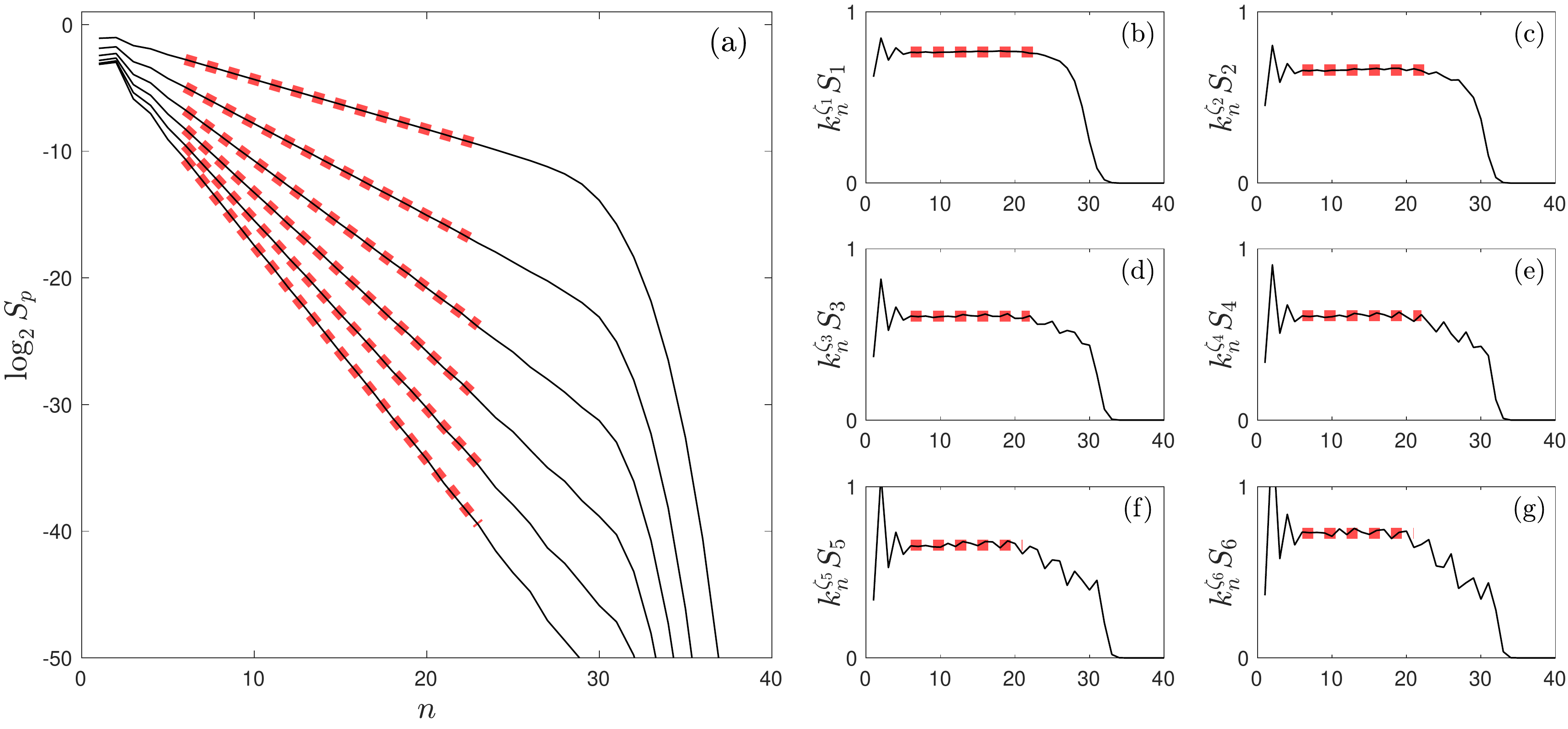}
\caption{(a) Logarithms of structure functions, $\log_2 S_p$, depending on shell numbers $n = \log_2 k_n$ for $p = 1,\ldots,6$ (from upper to lower graphs). Red dashed lines show the power-laws $\propto k_n^{-\zeta_p}$ in the inertial interval. (b-g) Compensated structure functions, $k_n^{\zeta_p} S_p$ for $p = 1,\ldots,6$ depending on the shell number $n$. Red dashed lines show constant prefactors corresponding to power-laws $\propto k_n^{-\zeta_p}$.}
\label{fig1}
\end{figure}

Figure~\ref{fig1}(a) shows the numerical results for structure functions (\ref{eq1_Sp}) corresponding to the orders $p = 1,\ldots,6$. With the linear interpolation (red dashed lines), estimated values of the anomalous exponents in (\ref{eq1_Sp2}) are 
    \begin{equation}
    \begin{array}{c}
	\zeta_1 = 0.3945 \pm 0.0008,\quad
	\zeta_2 = 0.7225 \pm 0.0018,\quad
	\zeta_3 = 1.0023 \pm 0.0019,\\[3pt]
	\zeta_4 = 1.2536 \pm 0.0032,\quad
	\zeta_5 = 1.4843 \pm 0.0051,\quad
	\zeta_6 = 1.6943 \pm 0.0042,
	\end{array}
    \label{eqNR_1}
    \end{equation}
which agree with the results obtained in~\cite{l1998improved} for a random large-scale forcing. Figures \ref{fig1}(b-g) show the compensated structure functions $k_n^{\zeta_p}S_p(k_n)$. They indicate that pre-factors in power-laws (\ref{eq1_Sp2}) remain constant up to small numerical fluctuations for the shells of inertial interval $6 \lesssim n \lesssim 22$. Notice that this interval is getting smaller for larger orders $p$. 

\subsection{Hidden scale self-similarity}\label{sec6A}

\begin{figure}[t]
\centering
\includegraphics[width=1.00\textwidth]{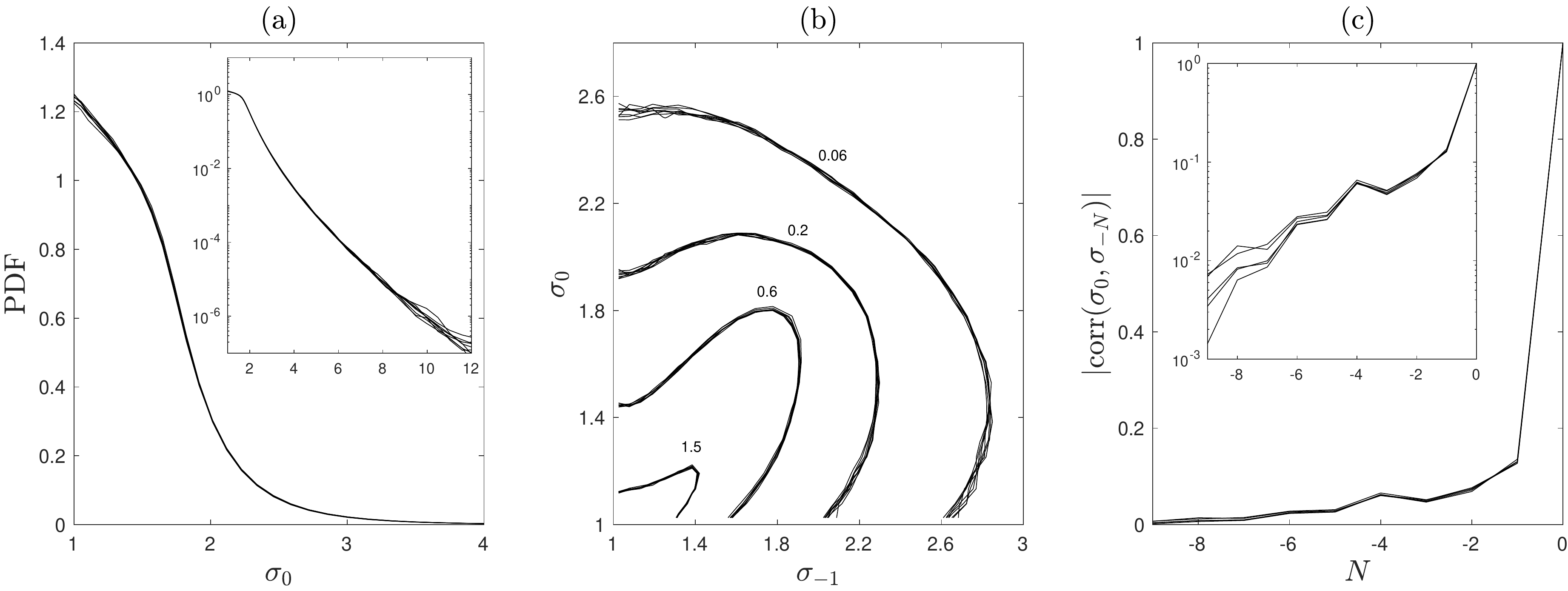}
\caption{(a) Probability density function (PDF) of the multiplier $\sigma_0$ computed for the reference shells $m = 11,\ldots,18$; inset shows the same functions but in logarithmic vertical scale. The accurate collapse of these graphs for different $m$ confirms the hidden scale self-similarity. (b) Level curves for the joint PDF of the two multipliers $\sigma_{-1}$ and $\sigma_0$. The results are shown for $m = 11,\ldots,18$ and the resulting graphs collapse to single curves up to small numerical fluctuations. (c) Absolute values of correlation coefficients between $\sigma_0$ and $\sigma_{N}$ depending on $N$. The graphs are shown for $m = 14,\ldots,18$ and demonstrate the locality property: decay of correlations among multipliers at distant shells.}
\label{fig2}
\end{figure}

Detailed analysis of the hidden self-similarity was done in the earlier work~\cite{mailybaev2021hidden}. Since, our analysis requires only the multipliers $\sigma_N$ defined in (\ref{eqAx_1mult}), we focus here on their statistical properties in the inertial interval. The hidden self-similarity means that the joint probability distribution of multipliers does not depend on the choice of the reference shell $m$. This property is confirmed in Fig.~\ref{fig2}, where the panel (a) shows the probability density functions (PDFs) of $\sigma_0$. These PDFs collapse into a single curve when computed for different reference shells $m = 11,\ldots,18$. These shells span roughly two and a half decades of wavenumbers from the central part of inertial interval. Similarly, the panel (b) demonstrates the hidden self-similarity in terms of the joint probability density of $\sigma_{-1}$ and $\sigma_0$. The panel (c) shows absolute values of correlation coefficients $\textrm{corr}(\sigma_0,\sigma_N)$ depending on the shells separation $N$ and computed for $m = 14,\ldots,18$. This figure verifies the locality of self-similar statistics: decay of correlations among multipliers at distant shells.

\subsection{Anomalous exponents in the self-similar statistics}\label{sec6b}

Next we consider measures $\mu_p^{(m)}$ defined in (\ref{eqZ_7prop}) and verify their self-similarity property (\ref{eqHSS_5}) with the Perron--Frobenius eigenvalues $R_p$ related by (\ref{eqHSS_8}) with the anomalous exponents. These measures will be analyzed  by using the corresponding density functions $f_p^{(m)}(\sigma_0,\sigma_{-1},\ldots,\sigma_{1-m})$. We express these functions from relation (\ref{eqZ_7prop}) as
    \begin{equation}
    f_p^{(m)}(\sigma_0,\sigma_{-1},\ldots,\sigma_{1-m}) = 
    k_m^{-p}\left\langle T_m^{-1} \right\rangle_t 
    \left(\prod_{J = 1-m}^{-1}\sigma_J^{p-1}\right) f^{(m)}(\sigma_0,\sigma_{-1},\ldots,\sigma_{1-m}),
    \label{eqNR_2}
    \end{equation}
where $f^{(m)}(\sigma_0,\sigma_{-1},\ldots,\sigma_{1-m})$ is a joint probability density function of multipliers for a specific choice of reference shell $m$. 

We define two types of convenient observables by integrating functions (\ref{eqNR_2}) with respect to all arguments but $\sigma_0$ or $(\sigma_0,\sigma_{-1})$, i.e.
    \begin{eqnarray}
    g_p^{(m)}(\sigma_0) &=& \int f_p^{(m)}(\sigma_0,\sigma_{-1},\ldots,\sigma_{1-m})\,d\sigma_{-1}\cdots d\sigma_{1-m},
    \label{eqNR_3}
    \\
    h_p^{(m)}(\sigma_0,\sigma_{-1}) &=& \int f_p^{(m)}(\sigma_0,\sigma_{-1},\ldots,\sigma_{1-m})\,d\sigma_{-2}\cdots d\sigma_{1-m}.
    \label{eqNR_4}
    \end{eqnarray}
Notice that the functions (\ref{eqNR_3}) and (\ref{eqNR_4}) represent marginal densities on the spaces of $\sigma_0$ and $(\sigma_0,\sigma_{-1})$, respectively. Hence, these functions can be computed numerically by using one- or two-dimensional histograms for the variables $\sigma_0$ and $(\sigma_0,\sigma_{-1})$: according to expression (\ref{eqNR_2}), the functions (\ref{eqNR_3}) and (\ref{eqNR_4}) are estimated within each histogram bin as sums of the quantities 
    \begin{equation}
    k_m^{-p}\left\langle T_m^{-1} \right\rangle_t 
    \left(\prod_{J = 1-m}^{-1}\sigma_J^{p-1}\right) \frac{\Delta\tau^{(m)}}{\tau_{\mathrm{tot}}^{(m)}\Delta_b}.
    \label{eqNR_2b}
    \end{equation}
Here $\Delta\tau^{(m)}$ is the segment of intrinsic time spent in a particular bin, $\tau_{\mathrm{tot}}^{(m)}$ is the total intrinsic time of the simulation, and $\Delta_b$ is the length (or area) of the bin.

\begin{figure}[tp]
\centering
\includegraphics[width=0.90\textwidth]{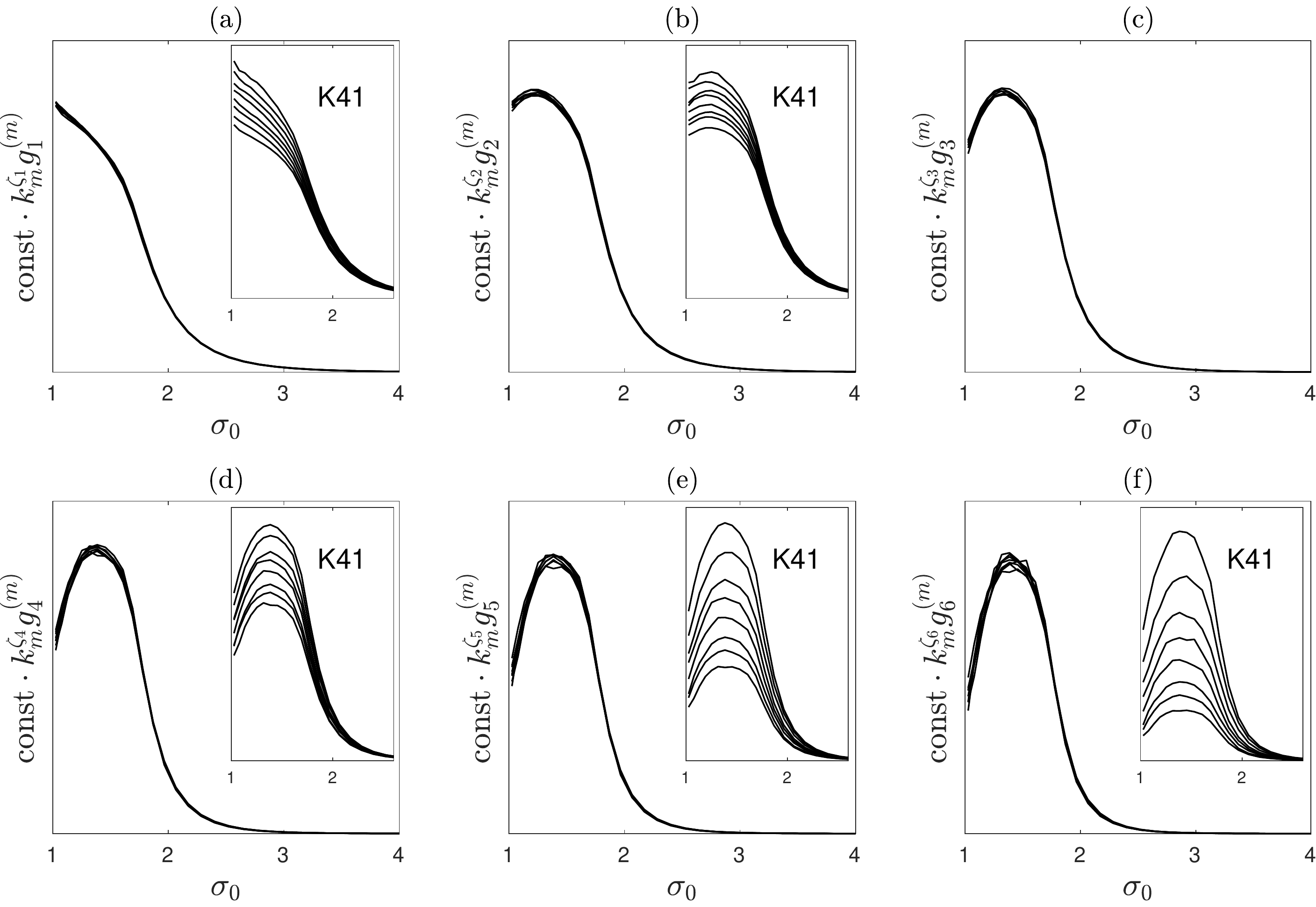}
\caption{Each panel shows compensated marginal densities $k_m^{\zeta_p} g_p^{(m)}(\sigma_0)$ for reference shells $m = 11,\ldots,18$ and fixed $p$. Panels (a-f) correspond to $p = 1,\ldots,6$. Collapse of the graphs for different $m$ verifies the relation between hidden self-similarity and anomalous scaling, as described by expressions (\ref{eqHSS_5}) and (\ref{eqNR_5a}). For contrast, we show in the insets the same functions but with $\zeta_p$ replaced by their K41 estimates $p/3$.}
\label{fig3}
\end{figure}

\begin{figure}[tp]
\centering
\includegraphics[width=0.85\textwidth]{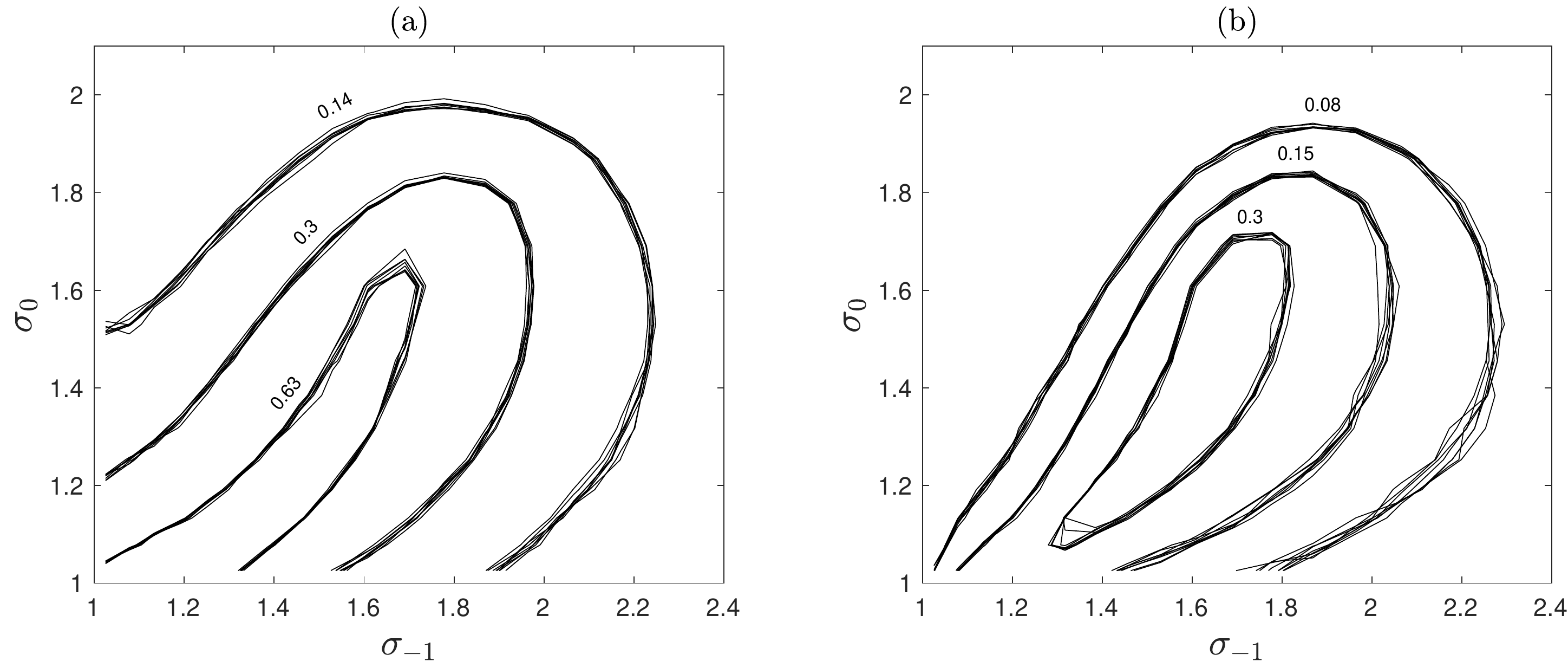}
\caption{Isolines of compensated marginal densities $k_m^{\zeta_p} h_p^{(m)}(\sigma_0,\sigma_{-1})$ for (a) $p = 2$ and (b) $p = 4$. Each isoline is presented by eight graphs, which correspond to reference shells $m = 11,\ldots,18$. Their collapse verifies the relation between hidden self-similarity and anomalous scaling, as described by expressions (\ref{eqHSS_5}) and (\ref{eqNR_5b}). }
\label{fig4}
\end{figure}

Let $R_p$ be the Perron--Frobenius eigenvalue with the corresponding eigenvector (measure) $\mu_p^{\infty}$ defined by the eigenvalue problem (\ref{eqHSS_4}). Similarly to (\ref{eqNR_3}) and (\ref{eqNR_4}), we denote the corresponding marginal densities by $g_p^{\infty}(\sigma_0)$ and $h_p^{\infty}(\sigma_0,\sigma_{-1})$. Our result in (\ref{eqHSS_5}) and (\ref{eqHSS_8}) implies the asymptotic relations
    \begin{eqnarray}
    g_p^{(m)}(\sigma_0) & \approx & c_p R_p^m g_p^{\infty}(\sigma_0) = c_p k_m^{-\zeta_p} g_p^{\infty}(\sigma_0),
    \label{eqNR_5a}
	\\[5pt]
    h_p^{(m)}(\sigma_0,\sigma_{-1}) & \approx & c_p R_p^m h_p^{\infty}(\sigma_0,\sigma_{-1}) = c_p k_m^{-\zeta_p} h_p^{\infty}(\sigma_0,\sigma_{-1}).
    \label{eqNR_5b}
    \end{eqnarray}
The scaling in these relations can be verified by showing that the functions 
    \begin{eqnarray}
    k_m^{\zeta_p} g_p^{(m)}(\sigma_0),\quad
    k_m^{\zeta_p} h_p^{(m)}(\sigma_0,\sigma_{-1}) 
    \label{eqNR_5c}
    \end{eqnarray}
with the anomalous exponents (\ref{eqNR_1}) do not depend on the reference shell $m$ within the inertial interval. Numerical graphs of functions $k_m^{\zeta_p} g_p^{(m)}(\sigma_0)$ indeed collapse as shown in Fig.~\ref{fig3} for $m = 11,\ldots,18$. In order to see the quality of this collapse, we present the same functions in the insets, but with the anomalous exponents $\zeta_p$ replaced by their K41 estimates $p/3$, demonstrating a clear dependence on $m$. Further tests are presented in Fig.~\ref{fig4}, where one can see a similar collapse for isolines of functions $k_m^{\zeta_p}h_p^{(m)}(\sigma_0,\sigma_{-1})$. Therefore, we verified numerically the asymptotic relation (\ref{eqHSS_5}), which derives the anomalous scaling from the hidden self-similarity.

\subsection{Anomalous exponents from Perron--Frobenius eigenvalues}

Finally, we verify relation (\ref{eqHSS_8}) deriving the anomalous exponents from the Perron--Frobenius eigenvalues of linear operators $\mathcal{L}_p^{\infty}$, as well as the functional form in (\ref{eqHSS_5}) given by the corresponding eigenvectors (measures) $\mu_p^\infty$.
For this purpose we must solve the eigenvalue problem (\ref{eqHSS_4}). Since the Perron--Frobenius eigenvalue $R_p$ is real and positive and has a largest absolute value, it can be computed by iterative methods, i.e., applying the linear operator $\mathcal{L}_p^{\infty}$ iteratively to an initial arbitrarily chosen measure. These iterations converge to the eigenvector of the dominant Perron--Frobenius eigenvalue. 

For estimating the operator $\mathcal{L}_p^{\infty}$ numerically, we use the decay of correlations among multipliers at distant shells. Namely, given $N > 0$, we approximate the conditional density of $\sigma_0$ as
    \begin{equation}
    \rho^{(m)} \approx \tilde{\rho}^{(m)}(\sigma_0|\sigma_{-1},\sigma_{-2},\ldots,\sigma_{-N}).
    \label{eqNR_6}
    \end{equation}
This approximation denoted with the tilde takes into account a fixed number $N$ of preceding multipliers $\sigma_{-1},\sigma_{-2},\ldots,\sigma_{-N}$ and neglecting the statistical dependence of $\sigma_0$ on the multipliers $\sigma_{-N-1},\sigma_{-N-2},\ldots$ at larger scales. 
Densities (\ref{eqNR_6}) do not depend on the reference shell $m$ in the inertial interval as a consequence of the hidden self-similarity, as we already established above in Section~\ref{sec6A}. 
Let us write expression (\ref{eqPF_4}) for the operator $\mathcal{L}_p^{(m+1)}$ in terms of densities $f_p^{(m)}$ and $f_p^{(m+1)}$ as
    \begin{equation}
    f_p^{(m+1)}(\sigma_0,\sigma_{-1},\ldots,\sigma_{-m}) 
    = \left( \frac{\sigma_{-1}}{\lambda}\right)^p
    \rho^{(m+1)}(\sigma_0|\sigma_{-1},\ldots,\sigma_{-m})
    f_p^{(m)}(\sigma_{-1},\ldots,\sigma_{-m}).
    \label{eqPF_4_dens}
    \end{equation}
Substituting approximation (\ref{eqNR_6}) for $m+1$ into equation (\ref{eqPF_4_dens}) and integrating with respect to $\sigma_{-N},\ldots,\sigma_{-m}$ yields  
    \begin{equation}
    \tilde{f}_p^{(m+1)}(\sigma_0,\ldots,\sigma_{-N+1}) 
    = \int \left( \frac{\sigma_{-1}}{\lambda}\right)^p
    \tilde{\rho}^{(m+1)}(\sigma_0|\sigma_{-1},\ldots,\sigma_{-N})
    \tilde{f}_p^{(m)}(\sigma_{-1},\ldots,\sigma_{-N}) 
    \, d\sigma_{-N},
    \label{eqNR_7}
    \end{equation}
where we introduced the partially integrated (marginal) densities
    \begin{equation}
    \tilde{f}_p^{(m)}(\sigma_0,\ldots,\sigma_{-N+1}) 
    = \int 
    f_p^{(m)}(\sigma_0,\ldots,\sigma_{1-m}) \prod_{J = 1-m}^{-N} d\sigma_J.
    \label{eqNR_8}
    \end{equation}
 
Relation (\ref{eqNR_7}) written in a compact form 
    \begin{equation}
    \tilde{f}_p^{(m+1)} = \tilde{\mathcal{L}}_p^{(m)}\left[\tilde{f}_p^{(m)}\right]
    \label{eqNR_7b}
    \end{equation}
defines a linear integral operator $\tilde{\mathcal{L}}_p^{(m)}$. For sufficiently large $N$, this operator $\tilde{\mathcal{L}}_p^{(m)}$ yields an accurate approximation of the operator $\mathcal{L}_p^{(m)}$ from (\ref{eqPF_4}) and (\ref{eqPF_3}) expressed in terms of densities. The latter approximates the limiting self-similar operator $\mathcal{L}_p^{\infty}$ in (\ref{eqHSS_2}) and (\ref{eqHSS_3}) for reference shells $m$ from the inertial interval. Therefore, we can compute the Perron--Frobenius eigenvalues $R_p$ numerically as dominant eigenvalues of operators $\tilde{\mathcal{L}}_p^{(m)}$, with the accuracy controlled by increasing the number $N$ of condition multipliers.

In our numerical procedure, we approximate the conditional probability density (\ref{eqNR_6}) using multi-dimensional histograms. Such analysis is limited because of high requirements for both the memory and computational resources, as well as by available statistics, which are all crucial for the design of a numerical method. For each $\sigma_J$, $J = 0,-1,-2,\ldots,-N$, we consider the interval $1 \le \sigma_J \le 2^{3.84} \approx 14.32$ covering all observed values; see the inset in Fig.~\ref{fig2}(a). This interval is partitioned into exponentially increasing bins. The binning is controlled by a single parameter $\Delta$ determining the smallest bin size $2^{\Delta}-1 \approx \Delta\log 2$. The bins are defined by setting their edges at $\sigma_J = 2^{s\Delta_J}$ with $s = 0,1,2,\ldots$, where 
    \begin{equation}
	\Delta_{0} = \Delta,\quad
	\Delta_{-1} = \Delta_{-2} = 2\Delta, \quad
	\Delta_{-3} = \Delta_{-4} = 4\Delta, \quad
	\Delta_{-5} = \Delta_{-6} = 8\Delta.
	\label{eqNR_7D}
    \end{equation}
The best approximation we could access numerically in (\ref{eqNR_6}) was $N = 6$ for $\Delta = 0.06$ and $N = 4$ for $\Delta = 0.03$. The chosen form of binning has two optimal properties: it uses smaller bins in the region of moderate values of $\sigma_J$ with larger probabilities, while larger bins capture rare large values of $\sigma_J$. Also, the choice (\ref{eqNR_7D}) assigns larger bins for larger $|J|$, which allows capturing the dependence on distant multipliers $\sigma_J$ within limited computational resources. The same bins are used for the numerical approximation of functions (\ref{eqNR_8}), and the integrals in (\ref{eqNR_7}) are computed as Riemann sums.

For computing the Perron--Frobenius eigenvalue $R_p$ and the corresponding eigenvector, we fix the reference shell $m = 18$ from the inertial interval, and select arbitrarily the initial function $\tilde{F}_{\mathrm{ini}}(\sigma_0,\ldots,\sigma_{-N+1})$. Then, we iterate the relation 
    \begin{equation}
	\tilde{F}_{i+1} = \tilde{\mathcal{L}}_p^{(m)}[\tilde{F}_i],\quad
	\tilde{F}_0 = \tilde{F}_{\mathrm{ini}}. 
	\label{eqNR_7it}
    \end{equation}
Before each iteration, the functions are renormalised to the unit $L_1$-norm: $\tilde{F}_{i+1} \mapsto \tilde{F}_{i+1}/\|\tilde{F}_{i+1}\|_{L_1}$. This iteration method converges (we used thirty iterations for a very accurate convergence) and yields the Perron--Frobenius eigenvalue $R_p$ with the corresponding eigenvector (density function) as
    \begin{equation}
	\label{eqNR_7conv}
	\frac{\|\tilde{F}_{i+1}\|_{L_1}}{\|\tilde{F}_i\|_{L_1}} \to R_p,\quad
	\tilde{F}_i \to \tilde{f}_p^{\infty}.
    \end{equation}
We control the accuracy of our numerical method by refining the bins (decreasing $\Delta$) and improving the statistical approximation (increasing $N$).

\begin{figure}[t]
\centering
\includegraphics[width=0.9\textwidth]{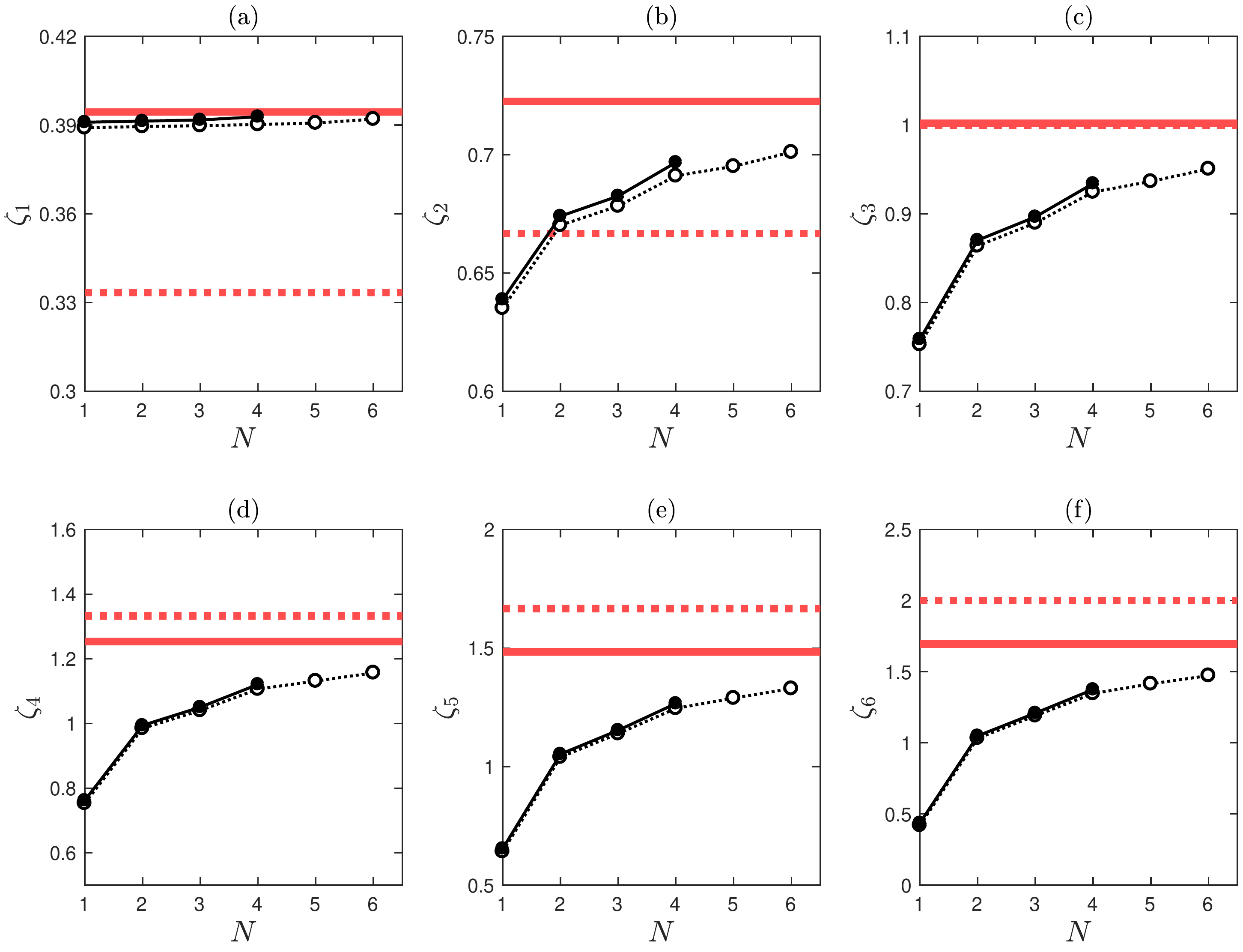}
\caption{Anomalous exponents $\zeta_p$ of orders $p = 1,\ldots,6$ (a--f) obtained in terms of the Perron--Frobenius eigenvalues (\ref{eqHSS_8}). The latter are computed numerically for different numbers $N = 1,\ldots,6$ in the approximation (\ref{eqNR_6}), and the eigenvalue problem is solved using the iterative method. All functions are estimated with multi-dimensional histograms: full circles connected by solid lines correspond to the binning parameter (size of smallest bin) $\Delta = 0.03$ and empty circles with dotted lines to $\Delta = 0.06$. Red horizontal lines denote the expected values of anomalous exponents from (\ref{eqNR_1}) and horizontal dotted lines indicate the K41 estimates $p/3$.}
\label{fig5}
\end{figure}
\begin{figure}[t]
\centering
\includegraphics[width=0.7\textwidth]{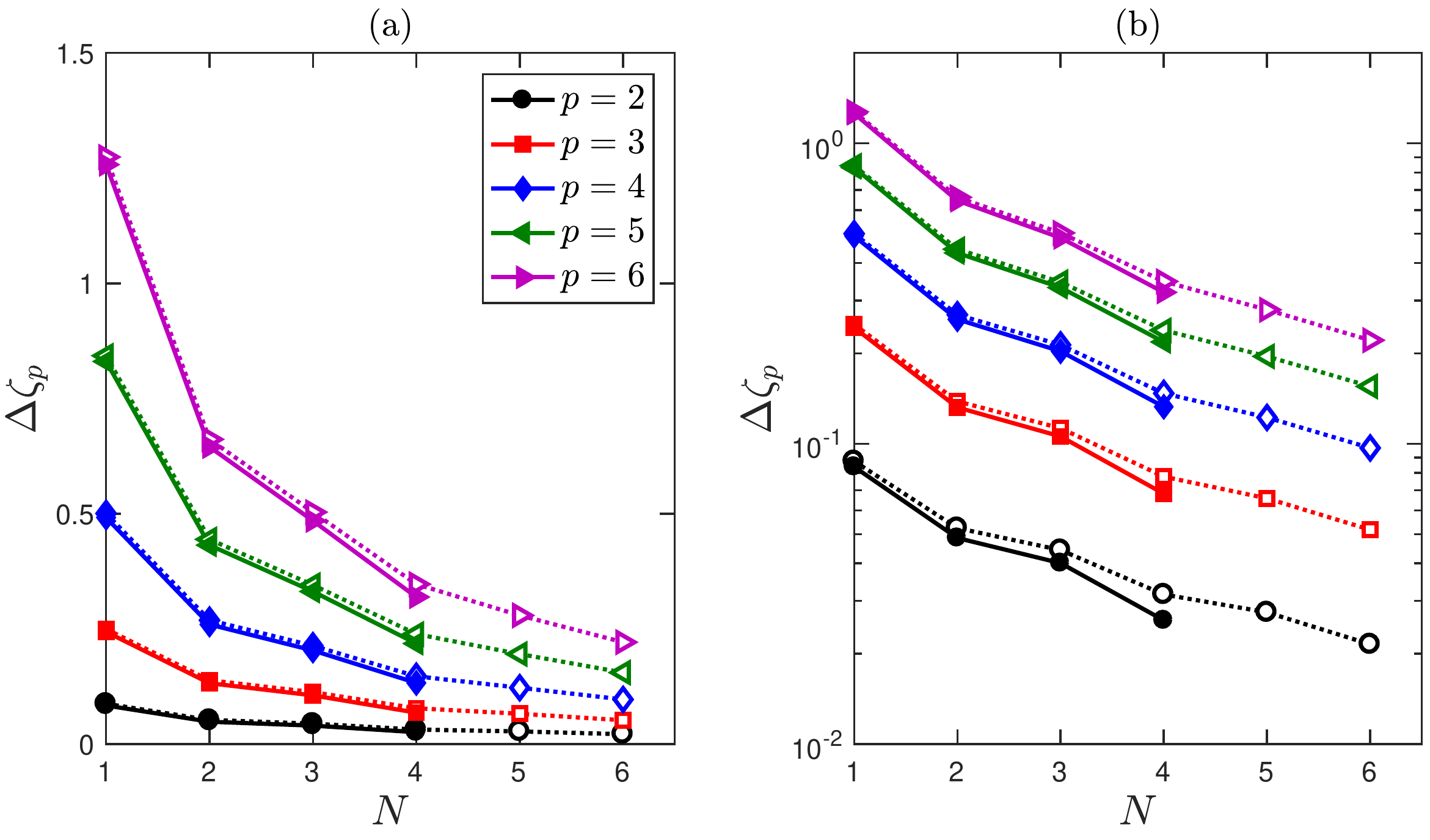}
\caption{Difference $\Delta\zeta_p$ between the values of $\zeta_p$ from (\ref{eqNR_1}) and the values of $\zeta_p$ computed in terms of  the Perron--Frobenius eigenvalues for different $N$; see Fig.~\ref{fig5}. The graphs are shown in (a) linear and (b) logarithmic vertical scales.}
\label{fig6}
\end{figure}

The proposed numerical scheme was implemented with different number of conditioned multipliers in (\ref{eqNR_6}) for the binning parameters (sizes of smallest bin) $\Delta = 0.03$ and $\Delta = 0.06$. The results are presented in Fig.~\ref{fig5}, where full circles connected by solid lines correspond to $\Delta = 0.03$ and empty circles connected by dotted lines to $\Delta = 0.06$. The statistical error of the results is small (does not exceed the size of plotted circles) as we verified by comparing with a simulation in a twice smaller time interval $0 \le t \le 1000$. Solid horizontal lines in  Fig.~\ref{fig5} mark the expected values (\ref{eqNR_1}), and dotted horizontal lines correspond to the K41 estimates $p/3$. We observe an excellent agreement for $\zeta_1$ in Fig.~\ref{fig5}(a) and  convergence to values (\ref{eqNR_1}) with increasing $N$ for the exponents $\zeta_2,\ldots,\zeta_6$ in Figs.~\ref{fig5}(b--f). The latter convergence is better seen in terms of differences $\Delta\zeta_p$ between exponents (\ref{eqNR_1}) and the estimates in terms of Perron--Frobenius eigenvalues for different $N$ and $\Delta$ as shown in Fig.~\ref{fig6}. The panel (b) of this figure shows the results in logarithmic vertical scale suggesting that all the exponents converge following the same pattern (supposedly, the same exponentially decaying mode) with the increase of $N$.

Recall that the exact value $\zeta_3 = 1$ of the third-order exponent is associated with the conservation of energy and very well agrees with the observations~\cite{frisch1999turbulence}; see (\ref{eqNR_1}). In our theory, we did not use the energy conservation and, therefore, the value $\zeta_3 = 1$ is not distinguished a priori. This is why we observe the gradual convergence in Figs.~\ref{fig5}(c) and \ref{fig6} for $\zeta_3$, just as for other higher-order exponents. Using relations (\ref{eq2_T}), (\ref{eq2}) and (\ref{eqAx_1}), one can express the energy $E = \frac{1}{2}\sum_n |u_n|^2$ in terms of multipliers, but this would give an expression with a sophisticated nonlinear dependence. It may be beneficial to explore this expression for understanding the interplay between the hidden symmetry and conservation laws and improving the numerical convergence; we leave such a study for a future work. 

\begin{figure}
\centering
\includegraphics[width=0.85\textwidth]{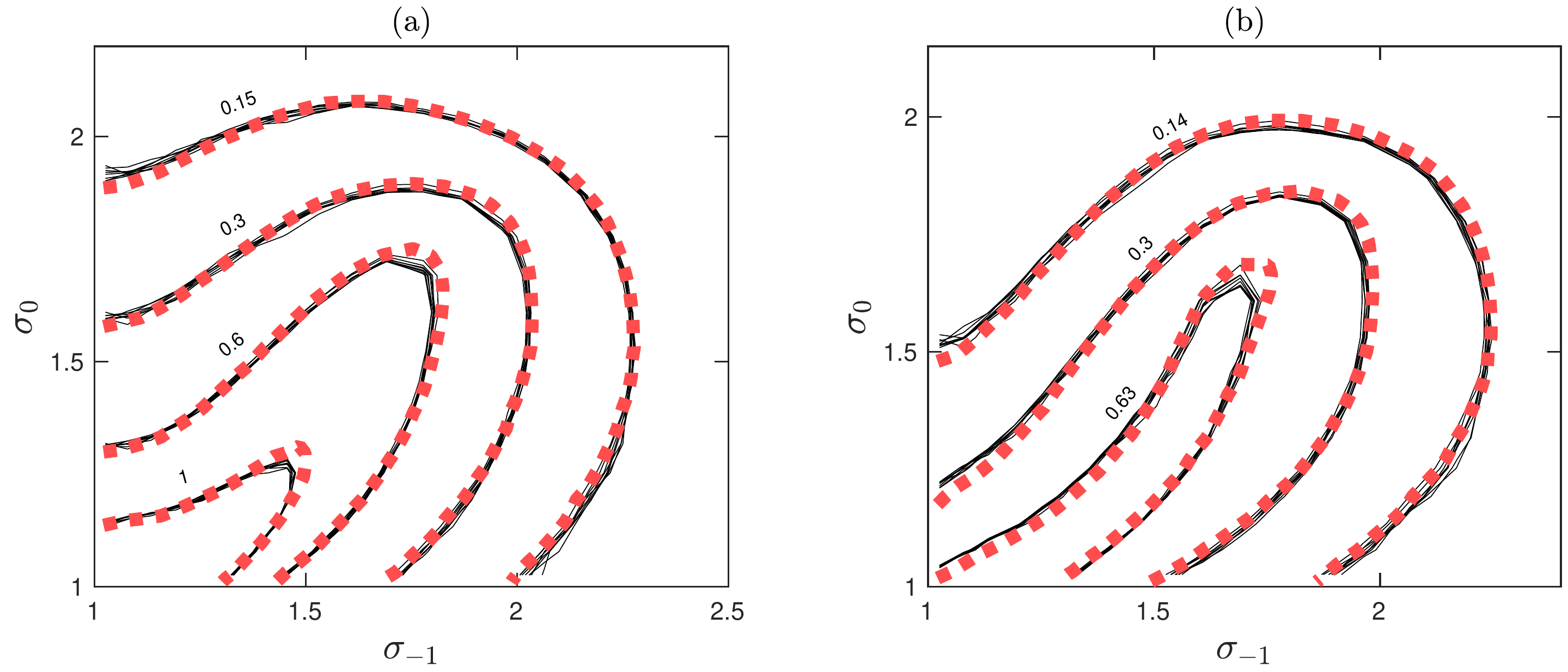}
\caption{Thin black curves are isolines of compensated marginal densities $k_m^{\zeta_p} h_p^{(m)}(\sigma_0,\sigma_{-1})$ for (a) $p = 1$ and (b) $p = 2$. Each isoline is presented by eight graphs, which correspond to reference shells $m = 11,\ldots,18$ as in Fig.~\ref{fig4}. Bold red dotted lines are the properly scaled functions $h_p^{\infty}(\sigma_0,\sigma_{-1})$ obtained in terms of Perron--Frobenius eigenvectors $\tilde{f}_p^{\infty}$ from (\ref{eqNR_7conv}). The agreement among all graphs verifies the asymptotic relation (\ref{eqNR_5b}) of the hidden self-similarity.}
\label{fig7}
\end{figure}

Another interesting comparison is presented in Fig.~\ref{fig7}. Here thin solid curves are isolines of the functions $h_p^{(m)}(\sigma_0,\sigma_{-1})$ for $m = 11,\ldots,18$. These functions were defined in (\ref{eqNR_4}) and their universality (independence of $m$) was established in Section~\ref{sec6b}; see Fig.~\ref{fig4}. According to the theory of hidden self-similarity formulated in expression (\ref{eqNR_5b}), these functions are equal (up to a constant factor) to the functions $h_p^{\infty}(\sigma_0,\sigma_{-1})$ computed in terms of Perron--Frobenius eigenvectors. Recall that while the former functions $h_p^{(m)}(\sigma_0,\sigma_{-1})$ are obtained directly from the simulation statistics as described in Section~\ref{sec6b}, the Perron--Frobenius eigenvectors are computed by the iterative method with a random initial density function; see (\ref{eqNR_7it}) and (\ref{eqNR_7conv}). The functions $h_p^{\infty}(\sigma_0,\sigma_{-1})$ computed numerically in terms of Perron--Frobenius eigenvectors with our finest approximation ($\Delta = 0.03$ and $N = 4$) are shown by bold dotted lines in Fig.~\ref{fig7} for $p = 1$ and $2$. An excellent agreement verifies our theoretical conclusions based on the hidden self-similarity.

\section{Discussion}\label{sec7}

We showed that intermittency in a shell model of turbulence is a direct consequence of hidden scale self-similarity. The hidden scaling symmetry appears when equations of motion are written in rescaled variables and times intrinsic to different scales of motion. Then, anomalous scaling of structure functions in the inertial interval is derived using the Perron--Frobenius eigenvalues of certain linear operators based on the self-similar statistics. We may conclude that the basic hypothesis of Kolmogorov that the scaling symmetry is restored statistically in the developed turbulence~\cite{frisch1999turbulence} is true for the shell model, however, not in the conventional sense. The key novelty is the choice of the symmetry: it is the hidden scaling symmetry which is relevant for turbulence rather than original scaling symmetries. 

We showed earlier within the general group-theoretical approach~\cite{mailybaev2020hidden} that the existence of hidden symmetries follows from non-commutativity of the temporal scaling (and also of the Galilean transform) with the evolution operator. Therefore, our approach can be extended to other systems, including the original Navier--Stokes equations~\cite{mailybaev2022hidden}. We know that the latter has the sweeping effect: large-scale motions affect the Eulerian small-scale statistics, and this feature is not captured by oversimplified shell models. However, the sweeping effect can be taken into account in the context of hidden symmetries as shown in~\cite{mailybaev2020hidden,mailybaev2022hidden}. 

We remark that chaotic dynamics is a complex phenomenon, and its understanding does not generally provide analytic procedures for computing all relevant quantities; one may recall the theory of Lorenz attractor as an example~\cite{tucker2002rigorous}. In our case, the hidden self-similarity explains the intermittent dynamics with anomalous scaling, but we need numerical simulations for computing accurate anomalous exponents. On the other hand, one can design specific models, where the analytic computation is carried out in full~\cite{mailybaev2021solvable}. Hence, for further development of the presented theory, more relevant studies may address  the robustness of hidden self-similarity, which ensures the importance of specific solvable examples, along with numerical investigations of the hidden self-similarity in real-world models. Another interesting direction that we did not address here is the relation of hidden symmetries with conservation laws like, e.g., the conservation of energy.

\section{Appendix}\label{sec8}

\subsection{Derivation of rescaled inviscid equations (\ref{eqZ_1}) and (\ref{eqZ_1b})}
\label{secA1}

Using (\ref{eq2}), we write 
    \begin{equation}
    \frac{dU_N}{d\tau} 
    = ik_mT_m^2\frac{du_{N+m}}{dt} 
    +ik_mT_mu_{N+m}\frac{dT_m}{dt}
    = ik_mT_m^2\frac{du_{N+m}}{dt} 
    +U_N\frac{dT_m}{dt}.
    \label{eqA1_1}
    \end{equation}
Substituting $T_m$ from (\ref{eq2_T}) yields
    \begin{equation}
    \frac{dU_N}{d\tau} 
    = ik_mT_m^2\frac{du_{N+m}}{dt} 
    +U_N T_m^3 \sum_{j = 0}^{m-1}k_j^2 \,\mathrm{Re} \left(-u_j^*\frac{du_j}{dt}\right).
    \label{eqA1_2}
    \end{equation}
Expressing derivatives using (\ref{eq1Euler}) and (\ref{eq2b}), we have 
    \begin{equation}
    \begin{array}{rcl}
    \displaystyle
    \frac{dU_N}{d\tau} 
    & = & \displaystyle
    -k_mT_m^2k_{N+m}\left(2u_{N+m+2}u_{N+m+1}^*
	-\frac{u_{N+m+1}u_{N+m-1}^*}{2} 
	+\frac{u_{N+m-1}u_{N+m-2}}{4} \right) \\[12pt]
    && \displaystyle
     +U_N T_m^3 \sum_{j = 1}^{m-1}k_j^3 \,\mathrm{Re} \left( -iu_j^*
     \left[2u_{j+2}u_{j+1}^*
	-\frac{u_{j+1}u_{j-1}^*}{2} 
	+\frac{u_{j-1}u_{j-2}}{4} \right]\right).
    \end{array}
    \label{eqA1_3}
    \end{equation}
It remains to write $k_mT_m^2k_{N+m} = k_N(k_mT_m)^2$ in the first term and $k_j^3 = k_{j-m}^3k_m^3$ in the second term. Then, using the second relation in (\ref{eq2}) one obtains equations (\ref{eqZ_1}) and (\ref{eqZ_1b}), where $J = j-m$.

\subsection{Explicit demonstration of hidden scale invariance}
\label{secA_2}

Let us explicitly demonstrate the invariance of system (\ref{eqZ_1_sym}) and (\ref{eqZ_1b}) with respect to the hidden scaling transformation (\ref{eqRV_plus}).
Using expressions (\ref{eqRV_plus}), we write
	\begin{equation}
	\label{eqA1}
	\frac{d\hat{U}_N}{d\hat{\tau}}
	= \frac{2}{\sqrt{1+|U_0|^2}} \,
	\frac{d}{d\tau} \frac{U_{N+1}}{\sqrt{1+|U_0|^2}}
	= \frac{2}{1+|U_0|^2} \,
	\frac{d U_{N+1}}{d\tau}
	-\frac{2U_{N+1}}{(1+|U_0|^2)^2}\,\mathrm{Re}\left(U_0^*\frac{dU_0}{d\tau}\right).
	\end{equation}
Substituting the derivatives from (\ref{eqZ_1_sym}) yields
	\begin{equation}
	\label{eqA2}
	\begin{array}{rcl}
	\displaystyle
	\frac{d\hat{U}_N}{d\hat{\tau}}
	&=& 
	\displaystyle
	\frac{2}{1+|U_0|^2} \,
	\left(-k_{N+1} B_{N+1}+U_{N+1}\sum_{J < 0} k_J^3\mathrm{Re}(U_J^*B_J)\right)
	\\[15pt] && \displaystyle
	-\frac{2U_{N+1}}{(1+|U_0|^2)^2}\,\left(-k_0\mathrm{Re}(U_0^*B_0)+|U_0|^2\sum_{J < 0} k_J^3\mathrm{Re}(U_J^*B_J)]\right).
	\end{array}
	\end{equation}
Manipulating the terms and recalling that $k_{N+1} = 2k_N$, $k_0 = 1$ and $k_J = 2k_{J-1}$, we have
	\begin{equation}
	\label{eqA3}
	\frac{d\hat{U}_N}{d\hat{\tau}}
	=
	-k_N\frac{4B_{N+1}}{1+|U_0|^2} 
	+\frac{16U_{N+1}}{(1+|U_0|^2)^2}\sum_{J \le 0} k_{J-1}^3\mathrm{Re}(U_J^*B_J),
	\end{equation}
where the sum is now taken over $J \le 0$. Using the last expression from (\ref{eqRV_plus}) twice for $N$ and $J-1$, we reduce Eq.~(\ref{eqA3}) to the form
	\begin{equation}
	\label{eqA4}
	\frac{d\hat{U}_N}{d\hat{\tau}}
	=
	-k_N\frac{4B_{N+1}}{1+|U_0|^2} 
	+\hat{U}_N \sum_{J \le 0} k_{J-1}^3\mathrm{Re}\left(\hat{U}_{J-1}^*\frac{4B_{J}}{1+|U_0|^2}\right).
	\end{equation}
After the substitution $J = J'+1$, this expression becomes (dropping primes)
	\begin{equation}
	\label{eqA5}
	\frac{d\hat{U}_N}{d\hat{\tau}}
	=
	-k_N\frac{4B_{N+1}}{1+|U_0|^2} 
	+\hat{U}_N \sum_{J < 0} k_J^3\mathrm{Re}\left(\hat{U}_J^*\frac{4B_{J+1}}{1+|U_0|^2}\right).
	\end{equation}
Using expression (\ref{eqZ_1b}), we have	
	\begin{equation}
	\label{eqA6}
	\begin{array}{rcl}
	\displaystyle
	\frac{4B_{N+1}}{1+|U_0|^2} 
	&=& \displaystyle
	\frac{4}{1+|U_0|^2}\left(2U_{N+3}U_{N+2}^*
	-\frac{U_{N+2}U_{N}^*}{2} 
	-\frac{U_{N}U_{N-1}}{4}\right) \\[15pt]
	&=& \displaystyle
	2\hat{U}_{N+2}\hat{U}_{N+1}^*
	-\frac{\hat{U}_{N+1}\hat{U}_{N-1}^*}{2} 
	-\frac{\hat{U}_{N-1}\hat{U}_{N-2}}{4} = \hat{B}_N,
	\end{array}
	\end{equation}
where we used the last expression from (\ref{eqRV_plus}) in the second equality. Similarly,
	\begin{equation}
	\label{eqA7}
	\frac{4B_{J+1}}{1+|U_0|^2} = \hat{B}_J.
	\end{equation}
Combining (\ref{eqA5})--(\ref{eqA7}), we obtain
	\begin{equation}
	\label{eqA8}
	\frac{d\hat{U}_N}{d\hat{\tau}}
	=
	-k_N\hat{B}_N
	+\hat{U}_N \sum_{J < 0} k_J^3\mathrm{Re}\left(\hat{U}_J^*\hat{B}_J\right).
	\end{equation}
Thus, we obtained the same equation (\ref{eqZ_1_sym}) but written in terms of the new variables and time, which proves that the transformation (\ref{eqRV_plus}) is the symmetry in the inertial interval.

\subsection{Fusion of space-time scaling symmetries}
\label{secA_2b}

Here we show that applying the rescaling procedure (\ref{eq2_T})--(\ref{eq2}) to the velocities $\hat{u}_n(\hat{t})$ from (\ref{eqSF1}) yields the hidden symmetry relations (\ref{eqRV_plus}).
First, using new shell velocities and time (\ref{eqSF1}), we express the new turn-over time (\ref{eq2_T}) as
    \begin{equation}
    \hat{T}_m(\hat{t}) 
    = \bigg(\sum_{j < m} k_j^2\left|\lambda^h u_{j+1}(t)\right|^2\bigg)^{-1/2}
    = \lambda^{1-h}\bigg(\sum_{j < m+1} k_j^2\left|u_j(t)\right|^2\bigg)^{-1/2}
    = \lambda^{1-h}T_{m+1}(t).
    \label{eqA2b_1}
    \end{equation}
Using (\ref{eqA2b_1}) with the temporal scaling relation $\hat{t} = \lambda^{1-h}t$ from (\ref{eqSF1}), we have
    \begin{equation}
    d\hat{\tau} = \frac{d \hat{t}}{\hat{T}_m( \hat{t})} = \frac{dt}{T_{m+1}(t)} 
    = \frac{T_m(t)}{T_{m+1}(t)} \frac{dt}{T_m(t)} = \frac{T_m(t)}{T_{m+1}(t)} \,d\tau,
    \label{eqA2b_2}
    \end{equation}
where we started with first relation in (\ref{eq2}) for the rescaled time $\hat{\tau}$ and then used it again for the time $\tau$.
Using (\ref{eq2_T}) and (\ref{eq2}), one obtains
    \begin{equation}
	\frac{T_m(t)}{T_{m+1}(t)} 
	=  \sqrt{\frac{\sum_{j \le m}k_j^2|u_j|^2}{\sum_{j < m}k_j^2|u_j|^2}}
	=  \sqrt{1+\frac{k_{m}^2|u_{m}|^2}{\sum_{j < m}k_j^2|u_j|^2}} 
	= \sqrt{1+k_{m}^2T_m^2|u_{m}|^2}
	= \sqrt{1+|U_0|^2}.
    \label{eqA2b_3}
    \end{equation}
Combining (\ref{eqA2b_2}) and (\ref{eqAx_1y}) yields the first hidden symmetry relation in (\ref{eqRV_plus}).

Similarly, expressing the new rescaled velocity $\hat{U}_N$ from (\ref{eq2}) and (\ref{eqSF1}), we have
    \begin{equation}
    \hat{U}_N = ik_m\hat{T}_m \hat{u}_{N+m} =  ik_m\lambda^{1-h}T_{m+1} \lambda^h u_{N+m+1} 
    = \frac{2T_{m+1}}{T_m}\, ik_mT_m u_{N+m+1} 
    = \frac{2U_{N+1}}{\sqrt{1+|U_0|^2}},
    \label{eqA2b_4}
    \end{equation}
where we used (\ref{eqA2b_1}), (\ref{eqA2b_3}) and again expression (\ref{eq2}) for $U_{N+1}$. This yields the second hidden symmetry relation in (\ref{eqRV_plus}).

\subsection{Proof of Theorem~\ref{theorem1}}
\label{secA_3}

First, let us reveal some basic properties of multipliers. 
Because of the boundary condition (\ref{eq2b}), expression (\ref{eq2_T}) yields
    \begin{equation}
	T_1(t) \equiv 1. 
    \label{eq2bb}
    \end{equation}
Using (\ref{eq2bb}) and (\ref{eqAx_1mult}) we have
    \begin{equation}
	 T_m = \prod_{J = 1-m}^{-1}\frac{1}{\sigma_J}.
    \label{eqAx_2a}
    \end{equation}
For $N = 0$, using (\ref{eqAx_1mult}) and (\ref{eqA2b_3}), one obtains
    \begin{equation}
	\sigma_0 
	= \sqrt{1+|U_0|^2}.
    \label{eqAx_1y}
    \end{equation}
The resulting expression can be inverted as
    \begin{equation}
	|U_0| = \sqrt{\sigma_0^2-1}.
    \label{eqAx_1inv}
    \end{equation}
Recal that hidden symmetry relations (\ref{eqRV_plus}) are equivalent to the change of reference shell $\hat{m} = m+1$. Thus, using (\ref{eqAx_1mult}) we see that multipliers change under hidden symmetry transformation (\ref{eqRV_plus}) as
    \begin{equation}
	\hat{\sigma}_N = \sigma_{N+1}.
    \label{eqAx_3}
    \end{equation}
In particular, from (\ref{eqM_1}) and (\ref{eqZ_8prop_B}) one has
    \begin{equation}
	\boldsymbol{\sigma} = \hat{\boldsymbol{\sigma}}_-.
    \label{eqAx_3vec}
    \end{equation}

For the reference shell $m$, it follows from (\ref{eq2}) that
    \begin{equation}
    \frac{1}{t}\int_0^t |u_m|^p dt' 
    = k_m^{-p}
    \left(\frac{1}{t} \int_0^{t} T_m^{-1} d t' \right)
    \left(\frac{1}{\tau}\int_0^{\tau} |U_0|^p T_m^{1-p} 
    d\tau' \right),
    \label{eqZ_3}
    \end{equation}
where we expressed $\tau = \int_0^t T_m^{-1}dt'$.
Taking the limit of large times and using expressions (\ref{eqAx_2a}) and (\ref{eqAx_1inv}),  structure functions (\ref{eq1_Sp}) are found as
    \begin{equation}
    S_p(k_m) = \lim_{t \to \infty} \frac{1}{t}\int_0^t |u_m|^p dt'
    = k_m^{-p}\left\langle T_m^{-1} \right\rangle_t
    \left\langle (\sigma_0^2-1)^{p/2} \prod_{J = 1-m}^{-1}\sigma_J^{p-1} \right\rangle_{\tau}.
    \label{eqZ_4}
    \end{equation}
Writing the last average in (\ref{eqZ_4}) in terms of probability measure from (\ref{eqAv_1}) yields expressions (\ref{eqZ_6prop}) and (\ref{eqZ_7prop}). 

Now let us consider the unit increase of the reference shell, $\hat{m} = m+1$, with the corresponding quantities denoted by the hats. Using (\ref{eqAv_1cond}) in expression (\ref{eqZ_7prop}) written for the new reference shell $\hat{m}$, we have
    \begin{equation}
    \begin{array}{rcl}
    	d\hat{\mu}_p(\hat{\boldsymbol{\sigma}}) & = & \displaystyle
    	k_{\hat{m}}^{-p}\left\langle T_{\hat{m}}^{-1} \right\rangle_t 
    	\left(\prod_{J = 1-\hat{m}}^{-1}\hat{\sigma}_J^{p-1}\right) d\hat{\mu}(\hat{\boldsymbol{\sigma}})\\[15pt]
	& = & \displaystyle
    	k_{\hat{m}}^{-p}\left\langle T_{\hat{m}}^{-1} \right\rangle_t 
    	\left(\hat{\sigma}_{-1}^{p-1}\prod_{J = 1-\hat{m}}^{-2}\hat{\sigma}_J^{p-1}\right) 
	\hat{\rho}(\hat{\sigma}_0|\hat{\boldsymbol{\sigma}}_-)
	\, d\hat{\sigma}_0 \, d\hat{\mu}_-(\hat{\boldsymbol{\sigma}}_-),
    \end{array}
    \label{eqZ_PR1}
    \end{equation}
where we also placed the factor $\hat{\sigma}_{-1}^{p-1}$ outside the product. We recast the resulting expression using $\hat{m} = m+1$, $k_{\hat{m}} = \lambda k_m$ and $\hat{\sigma}_J = \sigma_{J+1}$ from (\ref{eqAx_3}) as 
    \begin{equation}
    	d\hat{\mu}_p(\hat{\boldsymbol{\sigma}}) =
    	\left(\frac{\hat{\sigma}_{-1}}{\lambda}\right)^p k_m^{-p}\left\langle T_{m+1}^{-1} \right\rangle_t 
    	\left(\prod_{J = -m}^{-2}\sigma_{J+1}^{p-1}\right) 
	\hat{\rho}(\hat{\sigma}_0|\hat{\boldsymbol{\sigma}}_-)
	\, d\hat{\sigma}_0 \, \frac{d\hat{\mu}_-(\hat{\boldsymbol{\sigma}}_-)}{\hat{\sigma}_{-1}}.
    \label{eqZ_PR1_a}
    \end{equation}
After some additional manipulations and changing $J = J'-1$,  we obtain (dropping the primes)
    \begin{equation}
    	d\hat{\mu}_p(\hat{\boldsymbol{\sigma}}) = 
	\left(\frac{\hat{\sigma}_{-1}}{\lambda}\right)^p
    	\hat{\rho}(\hat{\sigma}_0|\hat{\boldsymbol{\sigma}}_-)
	\, d\hat{\sigma}_0
    	\left[ k_{m}^{-p}\left\langle T_{m}^{-1} \right\rangle_t 
    	\left(\prod_{J = 1-m}^{-1}\sigma_J^{p-1}\right) 
	\frac{\left\langle T_{m+1}^{-1} \right\rangle_t}{\left\langle T_{m}^{-1} \right\rangle_t}\,
	\,\frac{d\hat{\mu}_-(\hat{\boldsymbol{\sigma}}_-)}{\hat{\sigma}_{-1}} \right].
    \label{eqZ_PR1_b}
    \end{equation}
Expression (\ref{eqZ_8prop}) follows from (\ref{eqZ_PR1_b}), (\ref{eqZ_7prop}) and (\ref{eqAx_3vec}) if we prove that
    \begin{equation}
    	\frac{\left\langle T_{m+1}^{-1} \right\rangle_t}{\left\langle T_{m}^{-1} \right\rangle_t}\,
	\,\frac{d\hat{\mu}_-(\hat{\boldsymbol{\sigma}}_-)}{\hat{\sigma}_{-1}}
	= d\mu(\hat{\boldsymbol{\sigma}}_-).
    \label{eqZ_PR2}
    \end{equation}
	
For proving relation (\ref{eqZ_PR2}) let us consider an observable $\varphi(\boldsymbol{\sigma})$. Using the relations (\ref{eqAx_3vec}) and 
    \begin{equation}
	d\tau 
	= \frac{d\hat{\tau}}{\sqrt{1+|U_0|^2}} 
	= \frac{d\hat{\tau}}{\sigma_0} 
	= \frac{d\hat{\tau}}{\hat{\sigma}_{-1}} 
    \label{eqZ_9tau}
    \end{equation}
following from (\ref{eqRV_plus}), (\ref{eqAx_1y}) and (\ref{eqAx_3}), we have 
    	\begin{equation}
	\frac{1}{\tau}
	\int_0^{\tau} \varphi \left(\boldsymbol{\sigma}\right) \, d \tau'
   	= \frac{\hat{\tau}}{\tau} \frac{1}{\hat{\tau}}
	\int_0^{\hat{\tau}} 
	\frac{\varphi\left(\hat{\boldsymbol{\sigma}}_-\right)}{\hat{\sigma}_{-1}} 
	\,d\hat{\tau}'.
    \label{eqZ_9pr}
    \end{equation}
Using the first relation in (\ref{eq2}) for both $m$ and $\hat{m} = m+1$, the pre-factor is expressed as
    	\begin{equation}
	\frac{\hat{\tau}}{\tau}
	= \frac{t^{-1}\int_0^t T_{m+1}^{-1} d t'}{t^{-1}\int_0^t T_m^{-1}d t'}.
	\label{eqZ_9prB}
	\end{equation}
In the limit of large times, expressions (\ref{eqZ_9pr}) and (\ref{eqZ_9prB}) yield
    	\begin{equation}
	\langle \varphi(\boldsymbol{\sigma}) \rangle_{\tau}
   	= \frac{\langle T_{m+1}^{-1} \rangle_t}{\langle T_m^{-1}\rangle_t}
	\, \left\langle \frac{\varphi(\hat{\boldsymbol{\sigma}}_-)}{\hat{\sigma}_{-1}} \right\rangle_{\hat{\tau}},
    \label{eqZ_10pr}
    \end{equation}
which implies expression (\ref{eqZ_PR2}) by the assumed ergodic property (\ref{eqAv_1}) and relation (\ref{eqAx_3vec}).

\bibliographystyle{plain}
\bibliography{refs}

\end{document}